\documentstyle[prl,aps,epsfig,floats,times] {revtex}
\begin{document}
\draft

\twocolumn[\hsize\textwidth\columnwidth\hsize\csname %
@twocolumnfalse\endcsname

\title{Experimental Evidence for Electron Channelling in Fe/Au (100) Multilayers}
\author{D. T. Dekadjevi, P. A. Ryan and B. J. Hickey}
\address{Department of Physics and Astronomy\\ E. C. Stoner Laboratory, University
of Leeds, Leeds LS2 9JT, United Kingdom}
\author{B. D. Fulthorpe and B. K. Tanner}
\address{Department of Physics, University of Durham, South Road, Durham, DH1 3LE,UK}
\date{\today} \maketitle

\begin{abstract} We present transport and structural data from
epitaxial (100) and (111) Au/Fe multilayers grown by molecular
beam epitaxy. From their analysis, we conclude that an electron
channelling mechanism, due to strong specular reflection of the
minority spin carrier at the Au/Fe interfaces, is responsible for
the high conductivity in the (100) multilayers.

\end{abstract}
]

The theory of the giant magnetoresistance (GMR) in magnetic multilayers is still developing \cite{bozec,bauer} and
a number \cite{theory} of recent papers have interesting predictions about the nature of the electronic states in
multilayers. In particular, the band structure that gives rise to spin-dependent scattering is thought to be
responsible \cite{hood92} for the spin-dependent confinement of electrons to specific layers within the system. A
potential step can exist at the interface between the ferromagnetic layer and the non-magnetic layer.  This
potential in the hybridised band structure affects the transmission of electrons through the interface. It should
be possible, therefore, for electrons to scatter specularly from such an interface, and theoretical work has
predicted the effect on the transport\cite{butler96,zhang98,stiles96,dieny93,bulka,schep} and coupling properties
in superlattices\cite{mathon93,stiles93,bruno95}.

This type of specular reflection is quite distinct from that
explored experimentally by Egelhoff et al\cite{elgelhoff97},
Swagten et al\cite{swagten96} and Yuasa et al\cite{yuasa} where
electrons are specularly scattered from an insulator or noble
metal/vacuum interface. In true channelling the electron is
confined to a layer within the material as a result of the band
structure. States are completely reflecting if there are no states
in the other material with the same parallel momentum.
Additionally, at points of high symmetry, the effectiveness of
coupling between s-p states of the spacer and the d-states of the
ferromagnet become important and lead to high reflection
probabilities that are different for majority and minority
carriers.  When the magnetic moments in adjacent layers are
parallel, one spin can have a high specular reflection coefficient
at both interfaces and the spacer acts as an electron waveguide.
If this spacer has a significantly lower resistivity than the
magnetic layer, the channelled electrons will act as a low
resistance shunt and thus enhance the GMR. In the antiparallel
state, there is no confinement in the Au.

According to the theory of Stiles\cite{stiles96}, the Fe/Au (100)
system is a good candidate in which to observe electron
channelling. The probabilities for transmission {\it into} the Fe
from the Au are 67${\%}$ higher for majority carriers compared to
the minority carriers. For transmission {\it from} the Fe into the
Au, the majority carriers have a 21${\%}$ higher probability of
transmission. On the other hand, the minority electrons are
largely confined to each material. The orientation between the
successive magnetic moments in a multilayer greatly affects the
channelling. Only in the saturated state where all magnetic
moments are aligned can the channelling occur. Also, the
channelling is related to the symmetry of the Fermi surface and
therefore to the epitaxial relationship in the superlattice. Large
channelling effects are predicted for (100) Fe/Au whereas the
Fe/Au(111) system is not expected to exhibit channelling as the
band mis-match is so large\cite{stiles97}.

In this letter, we present experimental evidence for this mechanism in (100) oriented Fe/Au multilayers grown by
Molecular Beam Epitaxy (MBE)\cite{ourmbe}. The key experiments involve the study of the variation of conductivity
and structural perfection as a function of the thickness of the Au layers. {Fe/Au} multilayers consisting of 20
repeats of a [Fe(10\AA)/Au(X)] structure, where X represents a thickness between 5 and {70 \AA}, were grown on a
{50\AA} Au buffer layer. In-situ RHEED measurements and high resolution x-ray diffraction data from Bragg planes
parallel to the surface showed that multilayers grown on polished (100) faces of MgO (with a {10\AA} Fe seed
layer) were epitaxial with (100) orientation, while those grown on (11$\bar2$0) sapphire (with a {30\AA} Nb seed
layer) were (111) oriented. The existence of superlattice satellites up to the seventh order around the 200 and
111 x-ray Bragg peaks for the respective systems showed that the interfaces were extremely sharp. Analysis of the
diffraction peak widths in reciprocal space scans of {30\AA} bilayer samples gives a lower limit to the
out-of-plane sub-grain size of 150 $\pm$ {20\AA} and 570 $\pm$ {50\AA} in the (100) and (111) systems
respectively. In these samples, for both orientations, there are many interfaces contained within each sub-grain.
It is important for the transport data presented later,  to note that we find the out-of-plane structure for the
(111) samples to be of much higher quality than the (100) samples.

Oscillations, as a function of Au spacer thickness, in GMR (measured using a 4-probe DC method at 4.2K) and in the
remanent magnetisation were observed in the (100) orientation only. Our measured values of {15 $\pm$ 2\AA} and {7
$\pm$ 2\AA} for the period of the exchange coupling compares well with predicted values of {17.2\AA} and
{5.0\AA}\cite{bruno91} and those determined elsewhere\cite{shintaku93,unguris94}. However, the decay length of the
coupling strength, as a function of spacer, thickness was unusually long. Several theoretical papers have
predicted that the amplitude of exchange coupling is a function of the extent of electron
confinement\cite{mathon93,stiles93,bruno95}. There was no coupling observed in the (111) samples and a maximum GMR
of only 6${\%}$ was found compared to 40${\%}$ for the (100) samples with antiferromagnetic coupling.

The conductivity at 4.2K, measured in a saturating magnetic field of 1T, as a function of Au spacer thickness is
shown in Fig 1. The conductivity of a single bilayer of [{Au(1000\AA)/Fe(10\AA)}] has been measured as
$0.88\pm0.05(\mu \Omega cm)^{-1}$ in the (100) orientation and $0.73\pm0.04(\mu \Omega cm)^{-1}$ in the (111)
orientation. It is immediately apparent that the data from Fig 1 should saturate, for large Au thickness, at
roughly the same value, i.e. the conductivity of thick Au.  We have found that while the (111) data can be
explained in the Fuchs-Sondheimer model\cite{TFP}, the (100) data cannot, for any reasonable values of the Fe or
Au conductivities. If electrons were to traverse many layers without scattering, the conductivity would depend on
the average scattering properties of many Au and Fe layers and would not vary strongly with the thickness of the
Au(100) spacer. However, if electron channelling exists in the (100) Au spacers, the mean-free path of the
specularly reflected electrons may be long, leading to a high conductivity, without electrons crossing a large
number of layers. The number N, of specular reflections prior to diffusive scattering at the Fe/Au interface, is
determined by the potential step at the interface and does not change with Au-thickness. Provided the mean free
path in the Au layer, $\lambda_{Au}$, is greater than the thickness of the Au,  the mean free path is directly
proportional to the product of the number of specular reflections and the Au thickness: $\lambda_{Au} \propto N
t_{Au}$. The total conductivity, $\sigma$, measured in-plane is then the thickness-weighted sum of the
conductivities of the individual Fe and Au layers, namely,
\begin{equation}
\sigma=\frac {20 \zeta N t^2_{Au}} {t_{tot}} + \frac {20 t_{Fe}}
{t_{tot}} \sigma_{Fe}
\end{equation}
where t$_{Au}$ and t$_{Fe}$ are the thicknesses of Au and Fe
layers, t$_{tot}$ is the total multilayer thickness, $\sigma_{Fe}$
the conductivity of the Fe layers and $\zeta$ represents the usual
constants from the Drude formula. Solid lines in Fig. 1 represent
fits using equation 1 assuming that $\zeta$ is the same for all
samples, independent of Au thickness and multilayer orientation.
The only free parameter is the value of N and we find that the
ratio of specular reflection coefficients between orientations,
N$_{100}$~/~N$_{111}$ is 2.7 $\pm$  0.1 with N$_{100}$ equal to 9.

\begin{figure} \centerline { \epsfig{figure=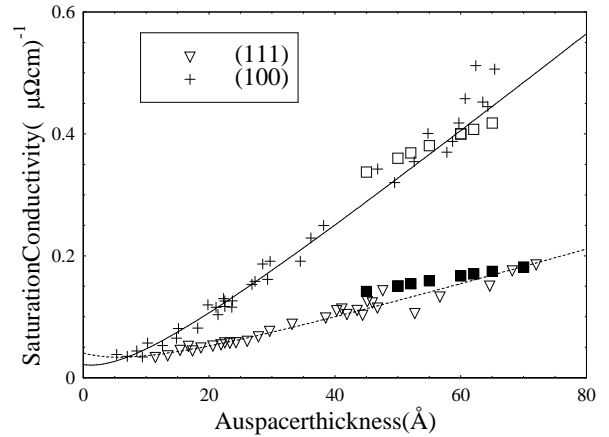,width=8cm}}
\caption{Saturated conductivity as a function of spacer thickness
for (100) (crosses) and (111) (triangles)
[Fe(10\AA)/Au(X)]$_{20}$ superlattices at 4.2K. Lines represent
fits of equation 1. Also shown is the saturation conductivity
calculated using the semi-classical model of Hood and Falicov for
(100) (open squares) and (111) (solid squares).} \label{Figure1}
\end{figure}

To go beyond this phenomenological model which assumes a spin
independent electron confinement, we have used the semi- classical
model proposed by Hood and Falicov\cite{hood92} to determine the
saturation conductivity in an infinite [Fe(10\AA)/Au(X\AA)]
multilayer as function of X.  We have calculated the conductivity
with and without specular reflections assuming spin dependent
scattering probabilities in the bulk and at the interfaces. The
spin dependent relaxation time and potential of the layer were
varied to keep the conductivity equal to that measured in the
bulk. At the interface, three possible cases are distinguished:
transmision (T), specular reflection (R) and diffuse scattering
(D) where T+R+D=1. The values of T and R are determined by the
layer potentials and D was set as 0.3 for all layers. We have set
the potential of the Au layer in agreement with the free electron
value of the Fermi energy. In order to confine the electrons, the
potential in the Fe must be varied. As expected, free electron
values of the Fermi energy for Fe do not work and therefore, to
introduce confinement in the Au, the potentials for majority and
minority carriers in Fe are changed.

Within the model we find that without electron confinement the
(100) Au conductivity must be ~25 times higher than the (111) Au
to explain the experimental data. This is unreasonable considering
the structural data and the measured conductivities. We therefore
fixed the conductivities of layers in each orientation to be equal
to the measured values and then switched on the channelling for
the (100) samples. The results are shown as squares in Fig 1.  The
agreement with the experimental data is very good when the
probability of specular reflection is set to 0.6 for minority
electrons and 0.3 for the majority electrons, which is close to
the values predicted by Stiles \cite{stiles96}. We also found that
varying D, the diffuse scatter at the interface, could not account
for the data. We conclude therefore, that channelling is required
to explain the observed differences in the conductivity of (100)
and (111) superlattices.

We have undertaken extensive structural measurements to demonstrate the validity of the above analysis which
assumes no significant change in the structural quality of the samples with Au thickness. In-plane diffraction
measurements have been carried in-situ using Reflection High Energy Electron Diffraction (RHEED) and ex- situ
using Grazing Incidence X-Ray Diffraction (GIXD). To perform an in-situ RHEED analysis, the growth of an {80\AA}
Au layer on {10\AA} of Fe has been interrupted at different stages. The patterns exhibit fine and continuous
streaks characteristic of a 2-D well ordered Au-surface, which is expected from the difference of free surface
energy of the materials. The full width at half maximum (FWHM) of the streaks is a measure of the in-plane
coherence length. As can be seen in Fig 2, this was found to be constant in both of the orientations for Au layers
thicker than {30\AA}, where the conductivity in Fig 1 shows significant dependence on the Au thickness. For thick
Au, the coherence length of the (111) samples was slightly larger than the (100).

\begin{figure} \centerline { \epsfig{figure=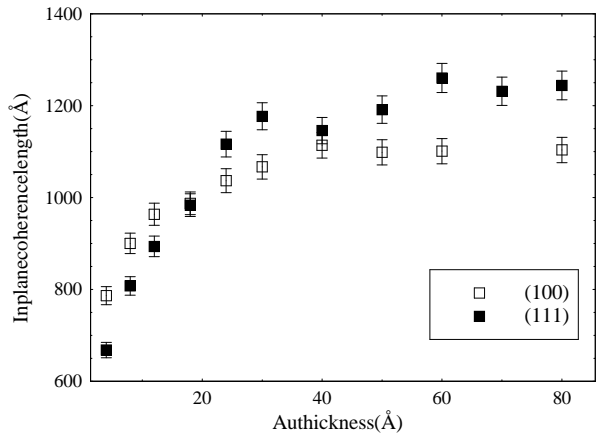,width=8cm}}
\caption{In-plane coherence length, determined from the width of
RHEED streaks, as function of the Au thickness and of the
orientation.} \label{Figure2}
\end{figure}

Grazing incidence x-ray diffraction experiments have been
performed on BM28, the XMaS CRG beamline, at the European
Synchrotron Radiation Facility in Grenoble. In the GIXD geometry,
the diffracting planes are perpendicular to the sample surface and
the diffraction peaks measured by rotation of sample and detector
about the normal to the surface. It is therefore sensitive to
in-plane structure and disorder in the Au and Fe layers. Use of a
grazing angle of the incident beam below the critical angle, (in
our experiments 0.2$^{\circ}$), limits the depth sensitivity to
approximately {30\AA}. The FWHM of the in plane $\theta $-2$\theta
$ scans are 0.8° for the (111) and 0.2° for the (100) orientation
samples respectively. Figure 3 shows scans of scattered intensity
as a function of the specimen angle about its surface normal taken
with the detector at the Au (022) reflection position. The 4- and
6-fold symmetries show the Au layers to be deposited with
well-defined (100) and (111) orientation in the MgO and sapphire
systems respectively. The FWHM of the peaks of Fig 3 provides a
direct measure of the mosaic width of in-plane crystalline
disorder. For the (100) samples, the mosaicity width was found to
vary from 0.49$^\circ$ to 0.47$^{\circ}$ for Au layer thicknesses
between {35\AA} and {70\AA} whereas for (111) the widths varied
from 2.35$^\circ$ to 2.02$^\circ$.  The absence of variation of
rocking curve FWHM with thickness in the (100) samples is strong
evidence against defect scattering being responsible for the
change in the conductivity\cite{modak}.

\begin{figure}\centerline { \epsfig{figure=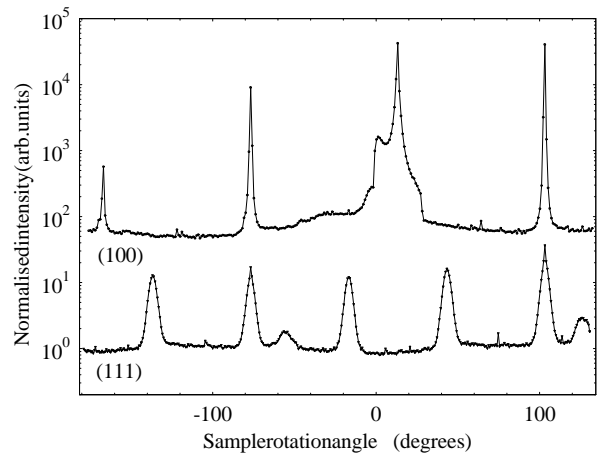,width=8cm}}
\caption{GIXD measurements of the in-plane rocking curves of the
Au 022 reflection for (a) the (100) and (b) the (111) orientation
multilayers.} \label{Figure3}
\end{figure}

The specular reflectivity data show Kiessig fringes and Bragg peaks for both (100) and (111) epilayers, and the
nominal thickness was found to be in good agreeement with that deduced from the x-ray results. Kiessig fringes and
Bragg peaks were also observed in the off-specular radial scans, indicating that a large proportion of the
interface roughness is correlated (conformal) through the multilayer. The thickness of individual layers, the
electron density gradient, conformal and random roughness, in-plane correlation length ($\xi$) and Hirst fractal
parameter, h, were obtained by matching the experimental data to that simulated for a fractal model structure
within the Distorted Wave Born Approximation\cite{bornwave,sinha}. Excellent agreement is found between
experimental and simulated data for transverse (q$_y$) scans, with good reproducibility between samples. For the
(111) Fe/Au epilayers grown on sapphire we find (Fig. 4a) that the interface roughness is highly correlated in
nature with a correlated-to-uncorrelated roughness ratio close to 16:1. The lateral correlation length, $\xi$, is
determined to be 250 $\pm$ {20\AA} with a fractal Hirst parameter h=0.20 $\pm$ 0.05. The (100) Fe/Au grown on MgO
also exhibits an interface roughness which is highly correlated in nature. Matching of simulated and experimental
data (Fig. 4b) yields a correlated-to-uncorrelated roughness ratio of 10:1 and a lateral correlation length $\xi$
=250 $\pm$ {20\AA}, with a fractal parameter h=0.28 $\pm$ 0.05. In all layers measured, the  r.m.s. amplitude of
the roughness in the (100) system is almost three times that of the (111) system, (9.3 $\pm$ {0.2\AA} compared
with 3.3 $\pm$ {0.2\AA}). The interface morphology is found to stay the same when varying the Au thickness.
Therefore, these structural properties imply that the (111) system should have the higher conductivity, contrary
to the observations.

\begin{figure}\centerline {\epsfig{figure=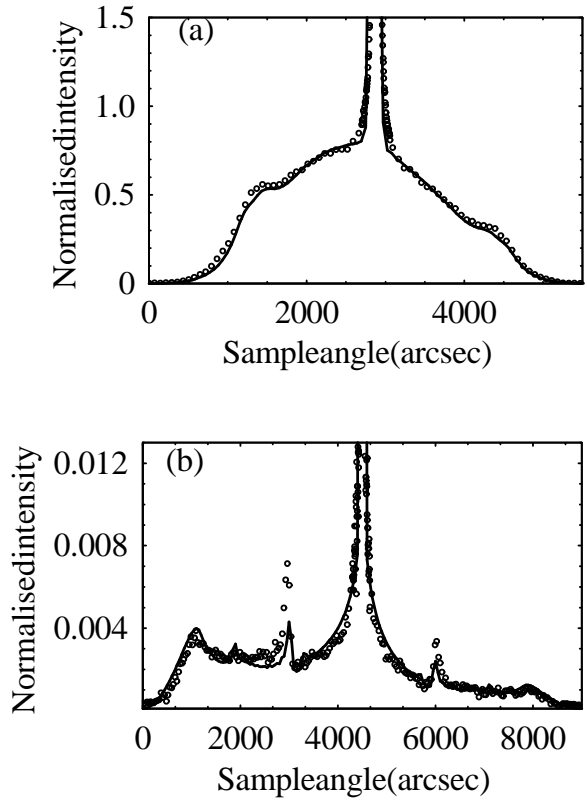,width=8cm}}
\caption{Grazing incidence x-ray transverse diffuse scan for
Fe/Au (100) on MgO (a) and (111) on sapphire (b) with an x-ray
wavelength $\lambda$=1.03\AA. Experimental data shown by points,
simulated fit by the solid line.  (The intensity differences
arise from the differing scattering angles in (a) and (b)).}
\label{Figure4}
\end{figure}

The values of $\sigma_{Fe}$, determined from the intercept in Fig
1, are very similar in the (111) and (100) materials. GIXD
measurements indicate that the (100) Fe is in good
crystallographic registration with the (100) Au as expected from
the lattice matching. In-situ RHEED data\cite{Dekadjevi} and ex
situ Medium Energy Ion Scattering\cite{Noakes98} experiments from
multilayers grown on sapphire showed that the growth of Fe (111)
on Au(111) is pseudomorphic for Fe thicknesses less than ~{10\AA}
in agreement with previous results \cite{Stroscio92}. The (111)
Fe/Au system eventually relaxes into its bulk structure, but only
after the Fe thickness becomes greater than 15\AA. For all samples
in this study, therefore, both orientations grow epitaxially.

 The structural studies show that neither the
variation of the saturation conductivity as a function of Au layer
thickness nor the difference between (100) and (111) oriented
multilayers can be attributed simply to the difference in sample
quality. We thus conclude that the data in Fig 1 demonstrate the
existence of electron channelling in Fe/Au (100) multilayers.\

Acknowledgements: We would like to thank M. Horlacher for coding
the Hood and Falicov model, J. Xu for some of the resistivity
measurements for the (111) samples, M. A. Howson and W. P. Pratt
for valuable discussions. The helpful and friendly assistance at
the ESRF of T.P.A Hase and XMaS beam line members S.D. Brown, D.F.
Paul, A. Stunault and P. Thompson and financial support from the
Engineering and Physical Science Research Council is gratefully
acknowledged.

\end{document}